# Using Machine Learning To Forecast Future Earnings

CUI Xinyue*

XU Zhaoyu*

ZHOU Yue*

# Table of Contents

# 1  INTRODUCTION

Earnings prediction has always been an important subject in accounting research in virtue of its proven relationship with market return (Beaver, 1968). Easton and Harris (1991) gives evidence that the cumulative abnormal return (CAR) is highly correlated with the EBIT, earnings per share, earnings yield, and other earnings indicators. Lipe (1987) and Clooins and Kothari (1989) also indicates the significance of the earnings response coefficient.

The opportunity to earn abnormal returns motivates researchers and practitioners to develop better earnings prediction methods for profit-making trading strategies. Questionnaires collected by Block (1999) proves accounting earnings are considered as the most important variable by analysts in securities valuation, considering that accurate earnings prediction would identify profitable investment opportunities from companies with remarkable performance.

The evolvement of earnings prediction not only reflects the development of accounting research but also employs the advancement of statistics and computer science subjects. Early researches relies on the random walk and time-series models to predict future earnings (Ball and Watts, 1972). Sooner had researches include more fundamental data in the prediction model based on linear regression (Deschamps & Mehta, 1980) or logistic regression (Ou and Penman, 1989).

Under the rapid development of computer science, numerous researches have highlighted the surprising potential of machine learning models, which could largely help to enhance accuracy on company earnings predictions and consequently earn abnormal profits. For example, Huang (2019) demonstrates in his paper that all three machine learning models adopted, Feed-forward Neural Network (FNN), Random Forest (RF), and Adaptive Neural Fuzzy Inference System (ANFIS) prevails over the market even without the specification in the case of sufficient data input. Meanwhile, Eakins and Stansell (2003) prove the neural network consistently outperforms throughout the full sample set referring to various market benchmarks. Inspired by the opportunity to automatically generate accurate earnings forecast, our research investigates a contemporary machine learning algorithm - LightGBM (a Gradient Boost Decision Tree model introduced by Microsoft in 2017) and its possible applications on the predictions of company fundamentals.

As indicated by Bishop (2006), machine learning algorithms heavily rely on samples (i.e., training data), including the LightGBM model. Hence, we cautiously select our samples from the quarterly reports of 3000 companies with the highest market capitalization in the US Equity Market. Past researchers have identified various fundamental factors contributing to earnings prediction, including currently observable financial ratios (Ou and Penman, 1989), univariate time-series earnings data (Griffin, 1977), and individual fundamental data, such as sales (Foster, 1977; Mabert & Radcliffe, 1974) and costs (Foster, 1977). In light of these established relationships, we select 73 most relevant fundamental variables from Compustat as our financial features (i.e., independent variables for prediction).

Additionally, past studies also attribute the variation of accounting earnings to macroeconomic variables (Brown and Ball, 1967; Gonedes, 1973; Basu, Markov, & Shivakumar, 2010) and contemporary abnormal returns (Abarbanell and Bushee, 1997), which encourage us to include six macroeconomic variables and stock return variables as well. The six macroeconomic variables reflect the Gross Domestic Product (GDP), inflation, market index, and interest rate levels during the reporting period.

To improve the comparability of our time-series and cross-sectional data, we selectively transfer the values of the above features into four different formats: year-over-year growth rate (YoY), quarter-over-quarter growth rate (QoQ), percentage of total assets (%Assets) and percentage of total revenue (%Revenue). On account of the availability of and correlation between these variables, we end up with 165 features in different formats, where 154 of them are derived from financial information on the company level. For these financial variables, we also adopt a 5-year (i.e., 20-quarter) look back period to incorporate the predictability of the time-series model, resulting in 3091 initial features in total.

To avoid the *curse of dimensionality* (Bellman, 1957), where Machine Learning models underperform the expectation due to the exorbitant degree of freedom and the difficulty of ordering distances in high-dimensional space, we first perform dimension reduction with Principal Component Analysis (PCA) technique that replaces the original features with their linear combinations that contain most of the initial information. This procedure effectively reduces the input dimensions from the original 3091 variables to a narrow range of 176 – 348 components, dependent on the difference between training samples and a designated fraction of variance explained by remaining components.

To fully discover the predictability of fundamental data on the company level, we select the LightGBM algorithm as our core machine learning classifier (i.e., tools for prediction). LightGBM is constructed under the classical gradient boosting framework using tree-based learning algorithms, but with faster training speed and higher efficiency compared with previous decision-tree models. This classifier has demonstrated noteworthy efficiency in other prediction tasks, including asset pricing (Ndikum, 2020) and cryptocurrency pricing (Sun, Liu, & Sima, 2018). These experiments also reveal the default model condition, known as hyperparameters, is not ideal in most cases (Ndikum, 2020). To relieve this research from extensive parameter tuning and configuration effort, we introduced Hyperopt, an optimization algorithm that can be used to automatically search for the best parameter combination within a pre-defined range efficiently, where we selected nine hyperparameters of LightGBM and their respective optimization ranges based on multiple trials to refine the model condition.

In terms of our dependent variable, considering the information limitation in our model (i.e., our model only contains public structured data), we transform the earnings prediction into classification problems to maximize the potential of the LightGBM algorithm. In addition to the sign of earnings change as estimated by Ou and Penman (1989), we also cut the relative earnings changes into three, six, or nine equal bins based on ranking, and test the performance of our model under all four scenario. The relative earnings changes are defined as the quarterly (QoQ) or annual (YoY) increase of earnings divided by total assets as at the last training quarter.

To ensure the stability and robustness of the prediction accuracy, we train and test our model on 40 overlapping and rolling subsets, which all consist of 80-quarter samples as the training and validation set for the LightGBM model and samples from the next quarter as the testing set, starting from 2008Q1. By moving testing set one quarter forward each time, we construct a 40-subset experiment where the last sample subset would train the model based on samples within 1998Q1 and 2017Q4 and calculate the out-of-sample accuracy with records from 2018Q1.

The out-of-sample accuracy rate would be compared with our two benchmarks, analysts' consensus estimations from I/B/E/S, and results of logistic regression (Ou & Penman, 1989; Hunt, Myers & Myers, 2019). To be specific, considering I/B/E/S analysts' consensus estimations target at Non-GAAP earnings, which is different from the original dependent variables we used (i.e., GAAP earnings from Compustat), we defined the benchmark accuracy of consensus estimations by using the same criteria for GAAP earnings to cut both consensus estimations and Non-GAAP earnings into multiple classes and calculate the ratio of the number of correctly classified samples among total samples accordingly.

Our 3-class prediction accuracy is respectively 52.1% and 52.7% on average for the relative earnings growth of the next year (YoY) and the next quarter (QoQ) over the 40 subsets. The QoQ prediction accuracy rate is also higher than the YoY prediction accuracy rate in 6-class and 9-class prediction. LightGBM shows the best performance in QoQ 3-class prediction, where its accuracy rate is 52.7%, compared with the 74.2% accuracy rate of analysts' consensus estimation. As referred to the concept of 'analysts' superiority' (Fried & Givoly, 1982; Das, Levine, & Sivaramakrishnan, 1998), given that analysts benefit from a broader information set, including non-accounting information and information released after the prior fiscal year or fiscal quarter, it is reasonable that our model underperforms the analysts' consensus estimations.

Furthermore, for the multi-class prediction, we test the accuracy rate in converging and diverging cases with analysts' consensus estimations. Converging cases show optimistic results that increase consensus prediction accuracy from 74.3% to 81.3% for QoQ 3-class prediction. This result indicates that our model could be used to verify the estimation from analysts and improve the probability of correct prediction in both YoY and QoQ cases.

While our model fails to reach the analysts' consensus estimation accuracy in multi-class prediction, the LightGBM model outperforms logistic regression models proposed in past literature for the sign of earnings changes prediction. Our model successfully improves the out-of-sample accuracy rate to 64.2% on average, compared with 62.0% of elastic net logistic regression (Hunt, Myers & Myers, 2019) and 63.1% of stepwise logistic regression (Ou & Penman, 1989).

Our model explores the sophisticated correlation between various variables and future earnings. LightGBM not only captures the non-linear or non-trivial reciprocities among financial, macroeconomic, and market variables but also considers the time-series effects. On the other hand, the repetitive experiments for 40 subsets and the introduction of the random split of validation sets during the training process ensure the robustness of our empirical results. This research demonstrates the huge potential of Machine Learning algorithms in the earnings forecast. The capacity to generate relatively accurate predictions for over a thousand companies within 20 minutes also casts some light on possible quantitative trading strategies through overcoming the information asymmetry in the market.

The remaining parts of this paper will start with a summary of the development of earnings prediction models based on past literature in section 2, followed by the elaboration of our experiment design and data processing of our model in section 3. Section 4 will discuss our results compared with analysts' consensus estimations and past literature as well as other supportive founding during the research. Additional information, including glossary and formula, can be found in Appendices for clarification.

## 2 LITERATURE REVIEW

This section will summarize various models that were used in previous earnings prediction papers. The profound interest expressed by previous researchers in company valuation and fundamental analysis is reviewed by Kothari (2001) in his capital markets research. He concluded that such kinds of accounting tests will be supportive in understanding capital market investment decisions and corporate finance. Therefore, numerous research papers emerged, which focus on conducting empirical experiments to predict earnings from fundamental data. Richardson, Tuna, and Wysocki (2010) summarized the most frequently cited research papers of earnings forecast which was published after 2000. The top ten cited papers apply a wide range of statistical models in terms of prediction techniques. For example, Lewellen (2004) adopts the ordinary least squares (OSL) regression on lagged dividend yield to predict monthly returns. Hong (2001) examines the abnormal accruals and other earnings components using iterative generalized nonlinear least squares estimation to forecast one-year-ahead earnings. Myers, Myers, and Skinner (2007) explore the momentum properties of company earnings, by computing the probability of random binomial variables through time-series models to predict split-adjusted EPS. A similar process of forecasting is adopted by Watts & Leftwich (1977), who tried the random-walk model and indicates its strong predictability power in the time-series behavior of company earnings. Furthermore, the researches done by Hayn (1995), Burgstahler & Dichev (1997), and Degeorge, Patel, & Zeckhauser (1999) document asymmetric patterns in the statistical distribution of earnings, which provides useful guidance for us to adopt a time-series embedded model in the process of prediction.

Due to the non-linear, dynamic variation of the financial and accounting information, it is impractical to use manual or traditional financial analysis to forecast earnings accurately. Therefore, a growing trend using machine learning techniques to replace and supplement current valuation models emerges. However, applying machine learning models on earnings prediction is far less popular than predicting stock returns, which leaves opportunities for us to explore this relatively unpopular field with the help of the rich techniques in financial market prediction. As stated by Henrique, Sobreiro & Kimura (2019), the highest cited papers regard three commonly used models, including support vector machines (SVMs) (Kim, 2003; Pai, Lin, 2005), neural networks (Campbell, 1987; Chang, Liu, Lin, Fan, & Ng, 2009; Chen, Leung, & Daouk, 2003), and autoregressive conditional heteroskedasticity (ARCH/GARCH) (i.e., a linear process to generate the values of the time series) (Engle, 1982; Bollerslev, 1986). The main path of predicting financial signals focused on the hybrid of different models, for example, the combination of ARIMA and SVMs by Pai & Lin (2005) and the combination of neural networks, which can generate momentous enhancements compared to linear combinations (Donaldson, Kamstra, 1999). The popularity and effectiveness of Neural Networks also extend to the field of predicting company earnings and give us remarkable results (Cao, Parry, 2009). Therefore, in this paper, we will grasp the opportunity to use popular machine learning models in earnings prediction and investigate the underlying financial substances from the result.

Among various machine learning models, we select the Gradient boosting decision tree due to its burgeoning popularity and compatibility. Jones, Johnstone, and Wilson (2015) provide some evidence that the generalized boosting, outperformed all other classifiers including conventional classifiers (e.g. LOGIT regression, PROBIT regression, and linear discriminant analysis) and fully nonlinear classifiers, (e.g. neural networks, support vector machines) in predicting the credit rating. Their prediction is derived from a set of financial indicators, market variables, analyst forecasts, and macro-economic variables, which is very similar to our model's input. Also, recent research conducted by Carmona, Climent, and Momparler (2019) using a gradient boosting approach to predict bank failure based on its high accuracy and predictive power. Also as stated by James, and Witten, Hastie, & Tibshirani (2017), the boosting models can process extreme value, nonlinearity, missing data, and categorical variables, which fits in the features of our input data. Therefore, in view of its potential of categorical forecast in the context of accounting and finance, we adopted one of the Gradient Boosting Decision Trees, LightGBM as our main predicting model. Also, to further increase accuracy, we adopt a principal component analysis (PCA) beforehand to reduce dimensions. The details of our model will be discussed in later sections 3.2.

## 3  DATA AND METHODOLOGY

### 3.1  Data

LightGBM requires larger sample size and lower dimension in order to generate a more robust prediction model (Yang & Zhang, 2018). Hence, to ensure the reliability and validity of our model, on the one hand, we maximize the size of company universe, and one the other hand, we limit input variables to the most useful factors proven by past literature and analysts' practice. The following sections will discuss our selection of the samples and different variables in detail. After pre-processing, we end up with 3091 features, including financial, macroeconomic, and market variables, and 161,753 samples.

#### 3.1.1  Samples

For a larger sample size, we initially selected the 3000 companies with the highest market capitalization over a 30-year period from 1988 to 2017. However, larger sample size may also introduce more noise affect the accuracy of our model due to accounting standards evolution over the past 30-year period and discretionary reporting formats allowed by GAAP (Zeff, 2015). As a result, we eliminate our preliminary selection to ensure the reliability and consistency of input samples.

First of all, to ascertain the accountability of samples from Compustat database, we eliminate companies without GVKEY assigned, which represent potential ETF, and we also eliminate companies with abnormal trading history, where its share price is below $1 during the corresponding sampling period. We also exclude companies from the utility (GICS:55) and finance (GICS:40) sectors due to their incompatible statement formats. Furthermore, considering the different fiscal year ends may create confusion to the machine, we align the timing of samples by using the calendar quarter and remove companies whose fiscal year-end has been changed or is not within March, June, September, and December. After elimination, there remain 4592 companies over the 30-year period as shown in Appendix 6. With GVKEY as identifications for each company, we downloaded 333325 samples for selected financial factors from 1983 to 2018, given our intended 5-year look back period of and the 1-year forecast period.

### 3.1.2 Financial Variables

Among all numeric variables from the Fundamentals Quarterly library of Compustat, we select 73 most relevant variables based on past literature. Both accounting importance and statistical significance with earnings have been taken into consideration during the selection. Particularly, Year-to-date variables have been converted into quarterly data to ensure an identical reporting period for all variables.

#### *3.1.2.1 Variables Selection*

Our variable selection adheres to the established statistical properties of accounting data from voluminous researches on the subject of earnings prediction. As examined by Wild and Halsey (2007), the published financial statement liberally implies the intrinsic value of a company. The comprehensive review by Hillmer, Larcker, & Schroeder (1983) presents the autocorrelation exists in earnings (Deschamps & Mehta, 1980; Brown & Rozeff, 1979; Griffin, 1977), sales (Foster, 1977; Mabert & Radcliffe, 1974), and costs (Foster, 1977). A survey screening on financial accounting and financial analysis textbooks (Ou and Penman, 1989) distinguishes 68 common variables. Lev and Thiagarajan (1993) extracted information from professional commentaries of the industry-leading publications and elect 12 financial factors that indicate the best representatives of earnings prediction factors. On the ground of Lev and Thiagarajan's research, Abarbanell & Bushee (1997) further underlined 9 out of 12 fundamental signals which were derived from analysts' reports and other financial statement analysis materials. Consolidating all works mentioned above, we select 73 financial variables (Appendix 4) from the balance sheet statement, income statement, and statement of cash flow.

Additionally, the effect of quarterly seasonality also drives the motivation to adopt the univariate time-series data (Griffin, 1977). Therefore, as emphasized by the numerous empirical researches, we look back five years for each financial variable to imply the time-series properties in our model. For example, besides the quarter net income for 1988Q1, we also include the quarter net income for the past 19 quarters from 1983Q1 to 1987Q4 to construct a 20-quarter (5-year) lagging period to avoid seasonality variation and include necessary trendy or drifting information (Deschamps & Mehta, 1980).

### 3.1.2.2  Variables Definition

To increase the cross-sectional comparability, we convert the original data into four different formats, year-over-year growth rate (YoY), quarter-over-quarter growth rate (QoQ), percentage of total assets (%Assets) and percentage of total revenue (%Revenue) selectively, except the quarterly total revenue (REVTQ) and quarterly total assets (ATQ) whose original data would also be included as the indication of the scale of companies. As disclosed in appendix 4, among the original 73 variables, 73 variables are converted into YoY format, 55 variables are converted into QoQ format, 38 variables are converted into %Assets format, and 33 variables are converted into %Revenue format. The selection is guided by the distribution of each variable, financial analysis techniques, and past literature.

In terms of YoY and QoQ format variables, according to Ou & Penman (1989), the percentage of incomes and income-related expenses have a higher significant level in predicting earnings. Hence, all income statement related variables have chosen both QoQ and YoY formats, while most balance sheet related variables are only included in YoY format. Similarly, in light of the frequent fluctuation of cash flow variables, it seems appropriate to include both QoQ and YoY formats to capture their instability.

For the %Revenue, and %Assets format variables, referring to the common size analysis technique, we use %Revenue format for all income statement and cash flow statement related variables and %Assets format for balance sheet related variables.

After conversion, to remove the outliers from the highly skewed distributions, the %Revenue and %Assets values are further transformed to their natural logarithm.

### 3.1.2.3  Data Cleansing - Outliers

We address the outlier issues by equalizing all negative values to zero and assigning maximum values for all 198 converted variables for the reason discussed below.

Due to the existence of negative value before conversion, the QoQ and YoY format variables are no longer faithful representations of the economic value of the original data. A case in point would be that for companies reaching $1 million net income from $0.1 million loss and $10 million loss, the growth rate of the former company would be higher by the formula employed for QoQ and YoY conversion. However, from the economic point of view, it is clear the latter company has made a greater breakthrough during the period. As a result, we forfeit the information from changes within negative domains and focus on the relationships between positive values by replacing all negative values to zero.

On the other hand, to clear the extreme values on the positive side, we assign the 95% percentile of converted values derived from positive original values as the maximum value for all converted variables. In this way, we eliminate the misleadingly high values due to division by zero.

### 3.1.2.4  Data Cleansing - Missing Values

Four methods have been used to deal with the massive amount of missing values arising from 201 variables after format conversion, including sample deletion, variable deletion, relevant fill-in, and constant fill-in.

Sample deletion is used on circumstances where there are missing values on crucial accounting variables during the current or look back period. The crucial accounting variables include Total Assets, Total Liabilities, Total Equity, Revenue, Net Incomes, and Cash & Cash Equivalent, which are compulsory disclosure items under *asset and liability view* required by FASB since the beginning (1983) of our sampling period (Zeff, 2015). The missing values here are a clear signal of dubious records.

Variable deletion is used on variables with over 70% missing rate. Although some high missing variables can be significant to earnings prediction, it would be a waste of resources to investigate repetitive missing values due to multiple formats. To illustrate, In-Process R&D (RDIPQ) has proven to be useful on future earnings forecast (Kothari, Laguerre, & Leone, 2002); however, both its YoY and QoQ format variables reveal over 95% missing rate, which can hardly be useful in the model construction. As a result, among the three formats used for In-Process R&D, we remove YoY and QoQ format variables, keeping %Assets format variable with the relatively lower missing rate to account for the relationship between In-Process R&D and future earnings.

Relevant fill-in means filling in with the rolling average over the past 1 to 20 periods on remaining missing values for %Assets and % Revenue formats. To decide the optimal fill-in method, we compare the average of square residuals for existing values assuming the application of filling in with different rolling periods. As shown in Appendix 7, for most variables, rolling average with one period, namely the forward fill, would minimize the average of square residuals and hence is the best choice. Meanwhile, to avoid indefinite fill-in into future samples, the relevant fill-in will only apply to every first eight missing values, provided that the current state of business can at most stayed the same for future two years.

For the rest missing values failed to be fixed with relevant fill-in or removed through variable deletion, we implement constant fill-in, which is simply fill-in with constant value -1 to imply the missing. The above procedures make sure the financial variables input into our model is clean and reasonable, thus improving the interpretability and usefulness of our model.

### 3.1.3 Macroeconomic Variables

The overall economy health and inflation rate can also vastly influenced future earnings (Brown and Ball, 1967; Gonedes, 1973; Basu, Markov, & Shivakumar, 2010). Consequently, after the investigation on the correlation between macroeconomic factors, we include 9 characteristic indicators as our macroeconomic variables from the quarterly observation of Thomson Reuters, including:

  i. YoY and QoQ of GDP
 ii. YoY and QoQ of personal consumption expenditure price index (PCE)
iii. YoY and QoQ of S&P 500 index

    iv. central bank market rate, short-term market rate (rM3), and long-term interest rate (rY10)

To be specific, in view of the time lag between fiscal period end and reporting date, for market-determined variables, including various interest rates and S&P 500 index, we deliberately use values from the next calendar quarter. The 90-day requirement imposed by SEC (SEC, 2003) ensure values from the next calendar quarter must be the result fully respond to the reports of the current sampling quarter.

### 3.1.4 Market Variables

Market prices as a partial reflection of the market expectation on future earnings (Abarbanell & Bushee, 1997) could also provide useful information about future actual earnings. As a consequence, we include the market returns of each company over the next four quarters (YoY) and the next one quarter (QoQ) against their respective S&P500 benchmarks as our market variables. Similar to macroeconomic variables, we also use the next quarter values for the market variables to include the market response towards the quarterly reports for the current sampling period.

### 3.1.5 Net Incomes and Analysts' Consensus Estimation

Our model is designed to predict the year-over-year (YoY) and quarter-over-quarter (QoQ) growth rate of the quarterly net incomes as a classification problem. As illustrated in below Figure 1, YoY is defined as the difference of net incomes between the next year and the past year divided by the total assets as at the current quarter $T_0$, while QoQ represents the change of net incomes during the next quarter instead.

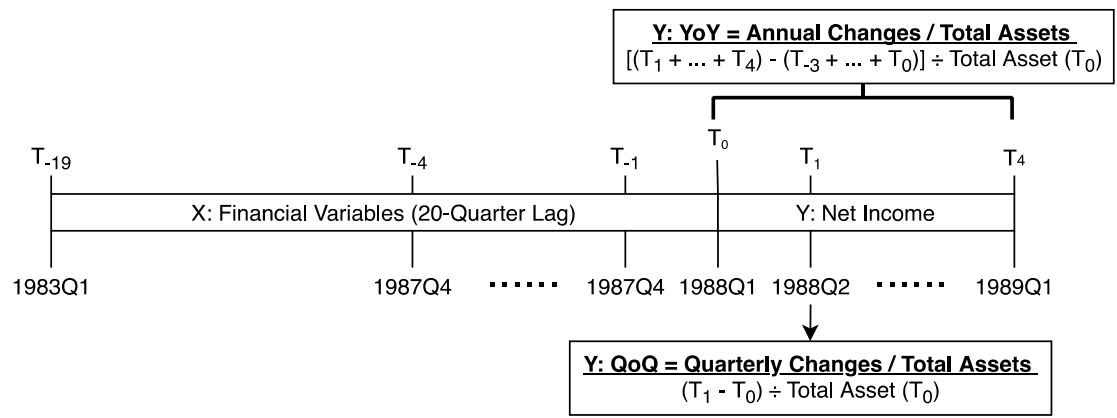

*Figure 1: Calculation of YoY and QoQ formats of net incomes*

Considering the noise in accounting and finance data, we choose to predict the relative profitability performance for each sample in the form of classification. For example, in a 3-class classification problem, by comparing this company's performance with other companies in the same period, we categorize its performance into three classes, including outperform(**2**), medium(**1**), and underperform(**0**). For instance, for the QoQ of all samples in 2008Q1, the top one third will be labelled as outperform(**2**), while the bottom one third will be labelled as underperform(**0**). The outperform label indicates the company had a relatively high net income growth compared to all peer companies used in 2008Q1 subset. Similarly, samples of YoY and QoQ format net incomes will also be cut into six and nine equal classes as well as two class based on their signs.

We use both mean and medium of analyst consensus estimations from I/B/E/S Summary database as our benchmark for evaluation. Given that analysts' consensus estimations target at Non-GAAP earnings (Gao & Liu, 2018), direct comparison between the consensus estimations and GAAP earnings from Compustat would distort the implied accuracy rate. As a result, to calculate the true analysts' accuracy for classification problems, we categorize both consensus estimations and Non-GAAP earnings into multiple classes with the same the same criteria for GAAP earnings, and then calculate the ratio of the number of correctly classified samples among total samples accordingly.

### 3.2   Experiment Design

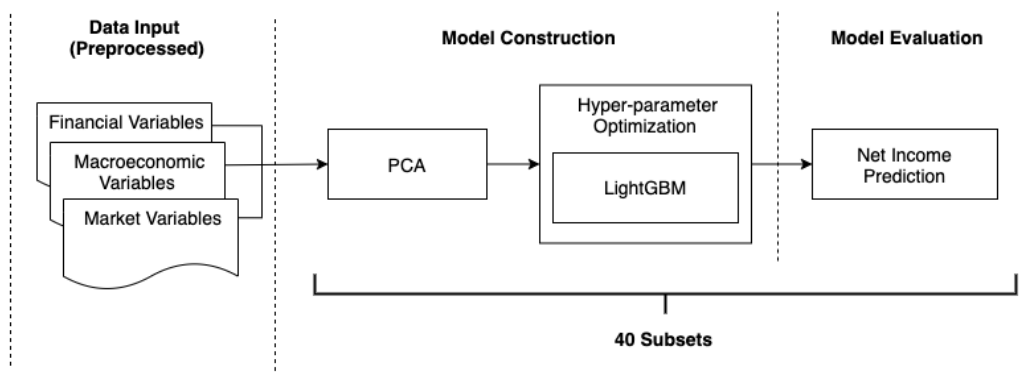

*Figure 2: Flowchart illustration of our experiment*

The overall focus of our experiment is to select, adjust and integrate LightGBM model with other ancillary state-of-the-art algorithms, which can predict the year-over-year (YoY) and quarter-over-quarter (QoQ) growth rate of earnings based on financial, macroeconomic, and market variables. As illustrated in Figure 2 above, our model construction has three major components:

    i.   Principal Component Analysis (PCA) for dimension reduction;

    ii.  LightGBM as the core machine learning classifier; and

    iii. Hyperopt to optimize the hyperparameters for LightGBM.

### 3.2.1   Dimension Reduction – PCA

In Machine Learning, many models may encounter several common obstacles mainly caused by sparse data samples and difficult distance ordering in high-dimensional situations, which are so called the "curse of dimensionality" (Bellman, 1957).

Specifically, in our project, initially there are 3091 different features (dimensions) after adding the artificial lagging. Thus, the degree of freedom would be so high that no way we could find sufficient number of samples to ascertain every aspect of them.

Moreover, there are also a lot of machine learning algorithm that involves the sequencing of "distance" (norm). When the number of dimensions goes up, the comparison of distance will be much more difficult, as the concept of "far" and "near" would become ambiguous in high-dimensional spaces.

One important way to alleviate such dimensional disaster is the *dimension reduction* technique, which refers to the mathematical transformations of the original high-dimensional attribute space into a low-dimensional subspace, where the sample density has been greatly increased, and the distance calculation has also become much easier. This process is intuitively illustrated in Figure 3. Considering our highly correlated initial features, it is likely that there has been an *embedded* low-dimensional subspace related to our learning task and easier to learnt by the machine learning model, LightGBM.

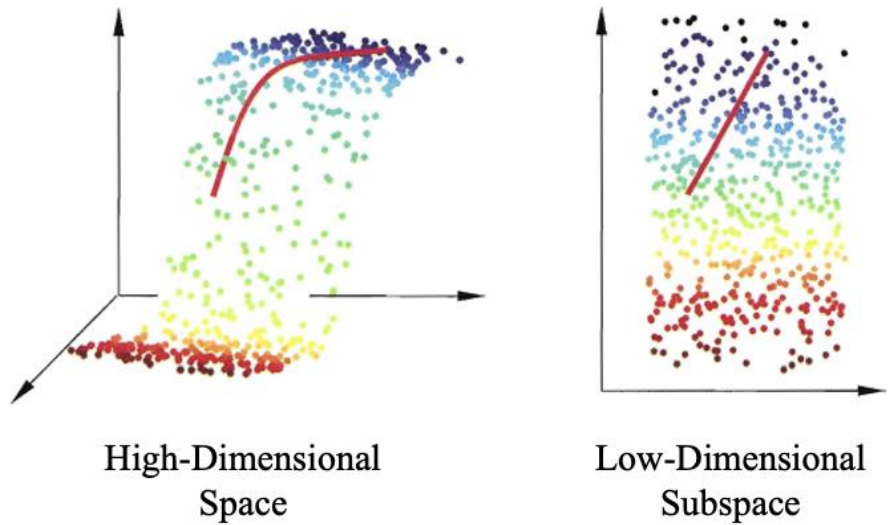

*Figure 3: Transformation from 3-D Samples to 2-D subspaces (Zhou, 2016)*

Among all the dimension reduction techniques, one of the most simple and intuitive methods would be the Ordinary Least Squares (OLS) regression, which selects the variables that have passed the t-test to directly get an effective subspace from the original high-dimensional sample space. Despite the better interpretability, this method may lack sufficient robustness due to collinearity or OLS inconsistency. Although using the pooled cross-sectional or panel regression may alleviate such conditions to some extent, it would replace what the Machine Learning techniques would do, making the data pre-processing overly complicated and becoming the actual model. As a result, we abandoned the attempt to explicitly identify the reduced dimensions and choose to use Principal Component Analysis (PCA) as a more robust dimension reduction technique.

Principal Component Analysis (PCA), on the other hand, investigates the linear correlations among samples with correlated variables, and effectively reduces the dimension by replacing a set of correlated variables into uncorrelated *components*. Each component is a combination of all or part of original correlated variables. The new reduced dimensions (i.e., components), could still preserve much of the variance dependent on the choice of user (Wold, Esbensen, & Geladi, 1987). Apart from the flexible choice of the number of dimensions at the expense of less variance explained, another benefit is the relatively fast and simple implementation through various python libraries, as well as better interpretability compared to more complex models such as autoencoder.

### 3.2.2 Machine Learning – LightGBM

The underlying concept of various machine learning algorithms is to use samples (i.e., training data) to generate a mathematical model (Bishop, 2006). Among these algorithms, the decision tree method has been proven to be more appropriate to predict in multiclass classification problems (i.e., categorizing the result of samples into multiple categories). The decision tree model has evolved from the initial single tree model to more complex for better performance and resource-saving, among which, two currently most widespread models are XGBoost and LightGBM.

For our specific case of the earnings prediction of specified corporations with selected variables, we will adopt one of the most powerful and effective Gradient Boosting Decision Tree (GBDT) models, LightGBM (Ke et al., 2017), which achieves great amelioration compared with traditional algorithms, including XGboost. Moreover, by evaluating the *feature importance* when constructing the decision tree, we can directly and intuitively learn which variables have more statistical significance for earnings prediction in our model. Particularly, the unique features of LightGBM include:

i. LightGBM can accelerate the training process and lower memory usage. This feature is based on the algorithm of histogram split, where it can divide continuous feature (i.e., attribute) values into discrete bins.

ii. LightGBM can be more precise using the leaf-wise (best-first) strategy to grow the decision tree (Figure 4), where most of the traditional boosting algorithms are using the level(depth)-wise method. When generating same number of leaves, the leaf-wise algorithm can incur much less information loss than the level-wise theory.

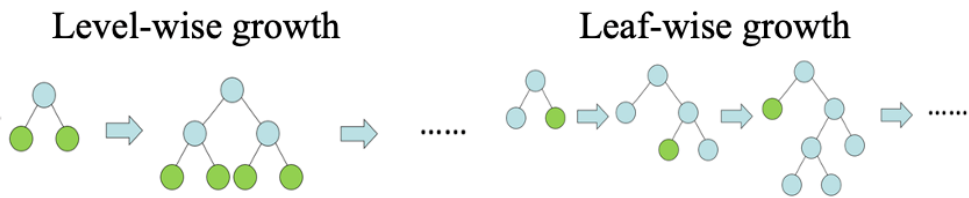

*Figure 4: Level-wise growth to Leaf-wise growth*

iii. LightGBM can further reduce the training time by its optimizations on parallel learning and collective communications, which also supports GPU for computations.

However, the leaf-wise algorithm may be easier to overfit, which means too focused on the training data and therefore hardly be reproducible on other cases (i.e., testing data) than the level-wise method especially when the training data are not sufficiently large as such algorithm can converge much faster. As a result, adjustments on parameters are of urgent needs, where we introduce several regularizations and restriction hyperparameters (Appendix 8) to effectively deal with such overfitting problems. Specifically, our model will utilize Hyperopt algorithm which tries and adjusts the hyper-parameters.

### 3.2.3 Hyperparameter Optimization – Hyperopt

Although the main idea of Machine Learning is to automatically generate a model without artificial interfere, Ndikum(2020) has demonstrated the default machine learning model condition, known as hyperparameters, is not far from optimal in financial problems. Hyperopt is one of the hyperparameters optimization algorithms sparing researchers from tedious manual tuning and configuration process. With a pre-defined range of hyperparameters (Appendix 8), Hyperopt benefits the training process of LightGBM model in terms of better accuracy and avoiding overfitting. We select the tuned hyperparameters based on the official including:

  i. *learning_rate*: the step length when we train or adjust our model. The smaller the value, the more precise LightGBM will be while approaching the optimized situation, and hence possibly reach better accuracy.
  ii. *max_bin*: the maximum number of bins (i.e., classes/categories) that feature values will be packed in. Smaller *max_bin* could accelerate the training process and avoid overfitting.
  iii. *num_leaves*: the number of leaves determining how large/deep/specified the decision tree will grow (be trained). As mentioned in the previous discussions, LightGBM adopts leaf-wise algorithm. Therefore, smaller *num_leaves* can be used to address overfitting problems.
  iv. *min_data_in_leaf* or *min_child_samples*: minimal number of records a leaf may have. If this parameter is set larger, it could prevent the training process from going too specified (i.e., overfitting). However, it may also cause under-fitting issues if not being appropriately set.
  v. *feature_fraction*, *bagging_fraction*, & *bagging_freq*: Parameters used to deal

with overfitting with smaller values. If *feature_fraction* is set smaller than 1.0 (i.e., the default value), LightGBM will randomly select this part of features for each iteration (i.e., training to grow the tree). Similarly, *bagging_fraction* determined the proportion of samples will be used for each iteration but without resampling, which is decided by *bagging_freq*, the parameter determining how frequent the 'bagging' operation will be conducted (i.e., bagging at how many rounds of iteration)

vi. *lambda_l1*, *lambda_l2*, & *min_gain_to_split*: Parameters related to regularization (i.e., the process of adding information in order to avoid overfitting with larger values). *lambda_l1* and *lambda_l2* corresponds to L1 & L2 Regularization Method respectively, while *min_gain_to_split* would limit the minimum information gain (i.e., improvement on the classification accuracy after each split) for each tree split in order to ensure that each further division is effective to prevent overfitting.

Hyperopt is specifically designed to automatically search for the best hyperparameters for specified Machine Learning models and tasks. Based on different hyperparameter combinations, the trained model with training set will be verified in validation set, which is split from the original training data for the purpose of obtaining the optimized parameter tuning result. The adjusted model will be used in the subsequent out-of-sample testing.

### 3.2.4  Implementation

We implement each round of test with the following procedures. Having successfully constructed the initial training set from the pre-processed data as explained in 3.1 section above, we would further adopt the dimension reduction techniques to avoid the curse of dimensionality. Considering high correlated variables may distort the result of PCA, the pre-processing process would take out variables with over 0.9 Pearson correlation with remaining variables as shown in Appendix 5.2, followed by the PCA to successfully reduce the original high-dimensional space to low-dimensional subspace with a explained variance ratio (i.e., remaining information) of 66% or 75%.

Afterward, we then adopted the *Hold-Out Validation Mechanism*, where each training set is constituted by samples of consecutive 80 quarters and the corresponding samples in the following quarter will act as the testing set. Moreover, the training set has also been further divided into two subordinate parts, *actual* training set and validation set to adjust the hyperparameters for better model performance with Hyperopt. To be specific, we test the performance of our model with difference validation set selection, in terms of size, from 1 to 20 quarters, and nature, using either the chronological last few or random few from the samples. Having obtained the best trail (i.e., the model with the highest accuracy on validation set), we shall then adopt such model to the testing set to calculate our out-of-sample accuracy.

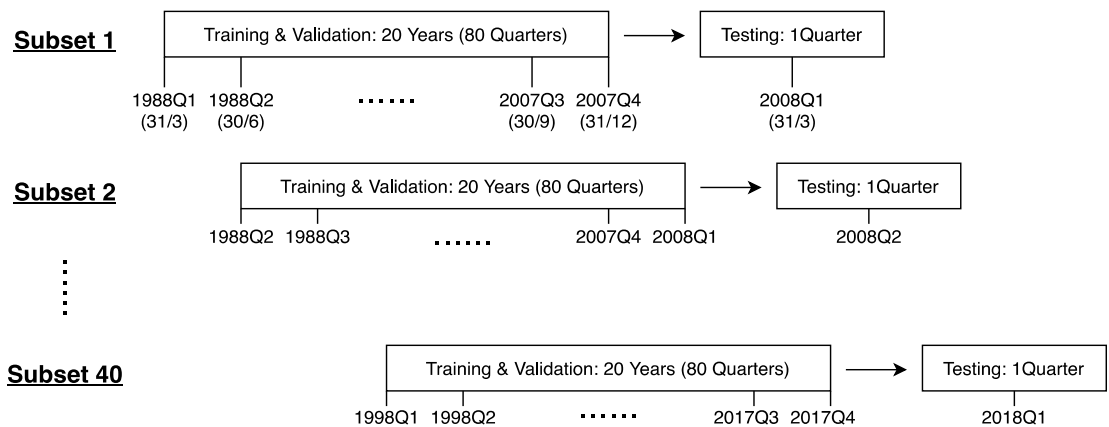

*Figure 5: Illustration of 40 rolling subsets selection*

Furthermore, to avoid the data snooping bias (Lo, 1994), we implement our model on 40 rolling and overlapping subsets over the 30-year sampling period (Figure 5), namely, 1988 to 2017, and expect to have meaningful and converging prediction results in each and every subset. The first subset would be samples from 1988Q1 to 2008Q1, where we use samples from the first 80 quarters (1988Q1 - 2007Q4) to build a model and test whether the model predicts correctly for samples from 2008Q1. Same procedures will be performed on other subsets with each quarter during the period of 2008Q1 to 2017Q4 as the quarter for out-of-sample prediction respectively.

## 4 RESULTS & DISCUSSION

### 4.1 PCA Results

Before conducting the training procedure of our model, we have firstly adopted PCA operation on all our 40 rolling subsets, where each of them has been set to generate the similar number of PCs for consistency. The relationship between the number of PCs and cumulative explained variance (i.e., the average ratio of all training sets) is shown in Figure 6 below. Due attention should be paid that we are supposed to manually set a threshold on the cumulative contribution rate for the principle components selected, as the number of dimensions required will correspondingly increase when such variance explained ratio is set higher. Having thoroughly considered such trade-off, we eventually decided to set 66% and 75% as the criteria for dimension reduction, resulting in between 176 and 348 Principal Components from the original 3091 dimensions.

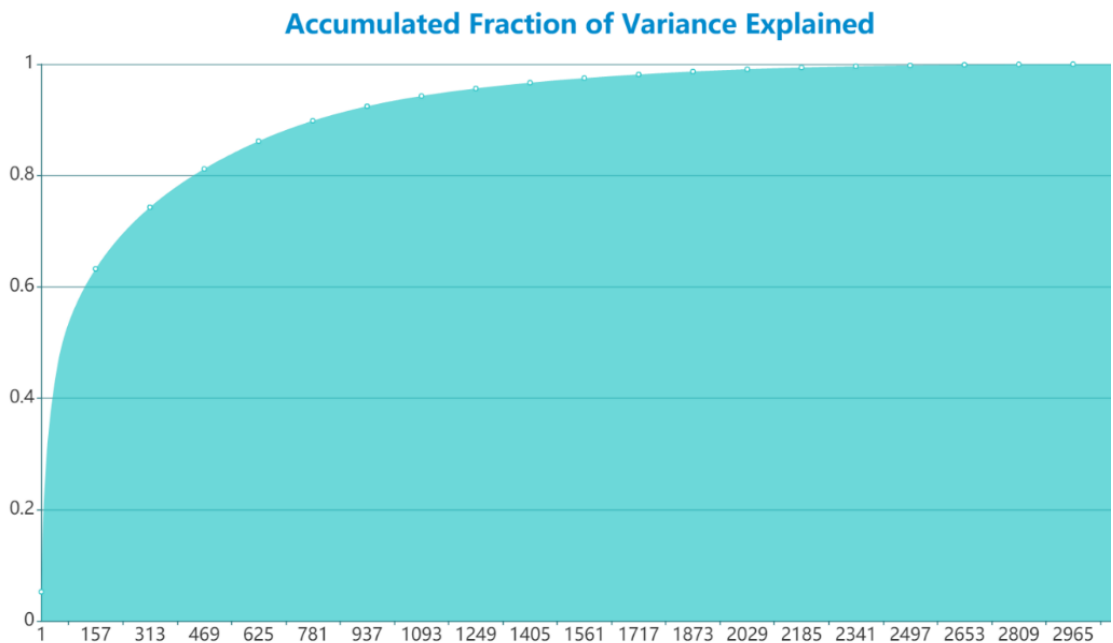

*Figure 6: Cumulative ratio of explained variance with respect to different numbers of principle components used*

## 4.2  LightGBM Results

### 4.2.1  Multi-class Prediction Results

As mentioned, we will use the reduced principle components for all qualified samples (i.e., X) to predict the corresponding relative earnings growth of following quarter (QoQ) or year (YoY) (i.e., Y) in terms of 3/6/9-class classification. Moreover, each training set has also been further divided into training and validation sets for both X and Y values for hyper-parameter optimization. The actual performance of such a trained and adjusted model will be tested on the corresponding testing set and the same procedure shall be repeated on all 40 subsets to ensure the robustness of our results.

The detailed chronological accuracy curve for all 6 conditions (3/6/9 class under YoY and QoQ) of our LightGBM model have been illustrated in the Appendix 9 and average results are summarized in Table 1 below, where the benchmark for such evaluation and comparison is the mean and median accuracy of consensus predictions made by several practitioners, which are obtained from the I/B/E/S dataset and are also correspondingly converted into the same number of categories for this classification problem.

*Table 1: Comparison on the accuracy performance between our LightGBM model and Consensus Predictions (Mean & Median)*

| | | Comparison of Multi-Class Prediction Performance on Average | | |
|---|---|---|---|---|
| Types of Depentent Variables | Number of Class | LightGBM | Conesnsus (Mean) | Conesnsus (Medium) |
| QoQ | 3 | *52.7%* | *74.3%* | *74.2%* |
| | 6 | 32.3% | 56.3% | 56.3% |
| | 9 | 23.4% | 46.2% | 46.3% |
| YoY | 3 | 52.1% | 71.6% | 72.0% |
| | 6 | 32.4% | 50.8% | 51.4% |
| | 9 | 23.5% | 40.4% | 40.7% |

It could be roughly identified from the results that there is still an arduous disparity on the prediction accuracy between our machine learning model and the works of analysts. However, it should also be well noted that such a phenomenon is not unexpected when comparing the consensus prediction with the results derived from statistical methods including time-series or others, which is so called analysts' superiority (Fried & Givoly, 1982; Das, Levine, & Sivaramakrishnan, 1998). Such situation can be attributed to multiple possible reasons and preliminarily, the real-life analysts may be largely benefited from an information advantage because analysts have access to a broader information set, which includes non-accounting information (i.e., information not disclosed to the public or information that can hardly be quantified as an input for the statistical/machine learning models) as well as information released after the prior fiscal year or quarter.

4.2.2 Conditional Accuracy Results

In addition, we have also conducted further researches on how the results of our LightGBM model can be actually utilized to advance the consensus predictions, where we calculated the conditional accuracy rate of prediction under the circumstances that the result of LightGBM converge (i.e., having the same forecast) or diverge with the consensus (Table 2). It is easily recognizable that when the consensus prediction has converged with our LightGBM results, the conditional accuracy rate is relatively higher than the normal average, which indicates that the LightGBM model could be used to further validate the results from analysts, thus to achieve better prediction performances.

*Table 2: Conditional accuracy when LightGBM converges & diverges with consensus prediction*

**Comparison of Converge Accuracy and Total Accuracy**

| Types of Dependent Variables | Number of Class | Converge | | Total | |
|---|---|---|---|---|---|
| | | LightGBM | Conesnsus (Mean) | LightGBM | Conesnsus (Mean) |
| QoQ | 3 | 68.4% | *81.3%* | 52.7% | *74.3%* |
| | 6 | 52.0% | 67.3% | 32.3% | 56.5% |
| | 9 | 43.3% | 58.4% | 23.4% | 46.4% |
| YoY | 3 | 63.0% | 78.4% | 52.1% | 71.6% |
| | 6 | 43.7% | 61.2% | 32.4% | 50.8% |
| | 9 | 33.5% | 52.1% | 23.5% | 40.4% |

### 4.2.3 Sign of Earnings Change Results

We have also compared our machine learning model with other traditional statistical methods (i.e., regression models) mentioned in previous literature, where the baseline task is to conduct the prediction of the signs of changes (i.e., increase or decrease) for the future earnings of specified corporations. Ou and Penman (1989) investigates the the probability of earnings increasing in the subsequent year by LOGIT model based on the 68 accounting descriptors, which is also employed by our model. Hunt, Myers, & Myers (2019) modifies the stepwise logit regression as well as elastic net logistic regression in order to predict the sign of future earnings changes based on a similar predictor base as ours. Therefore, the results generated by our model are compatible to compare with theirs. The comparison results are shown below (Table 3), which manifests that our model has surpassed the Stepwise Logistic Regression method and Elastic Net Logistic Regression method that have been widely adopted in the current industry.

*Table 3: Comparison on the accuracy performance between our LightGBM model and previous statistical models*

**Comparison of Prediction Performance of Sign of Earnings Changes**

| Methods | Our Paper | Ou and Penman (1989)* | Hunt, Myers & Myers (2019)** |
|---|---|---|---|
| LightGBM | 64.2% | | |
| Stepwise Logistic Regression | | 63.1% | 62.1% |
| Elastic Net Logistic Regression | | | 62.0% |

* This paper predicts probablity of earnings increase in the subsequent year. Here shows the average accuracy.

** Here presents the maximum accuracy in both case.

### 4.2.4 Feature Importance Analysis

Additionally, in virtue of the high interpretability of machine learning models, we have also conducted a truly innovative approach to decompose the five features that have the largest *Feature Importance* when being used to predict the earnings in our LightGBM model, which could help us to further obtain the ten most original variables that have the largest contributions to such forecast, thus to provide auxiliary insights to the analysts on which items in the financial database should they pay more attention to when predicting the future earnings.

Appendix 10 demonstrates the feature importance analysis based on the latest training-testing subset for the QoQ 3-class prediction model, where the out-of-sample accuracy rate amounts to 48.7%. We count the occurrence of the top ten important variables for the top five important components after PCA. In this case, all 50 variables are financial variables, where the three most critical variables are respectively total revenue(REVTQ), total operating expense(XOPRQ), and income taxes payable(TXPQ). Moreover, we also summarize the 50 important financial variables by the lagging period and converted formats, which shows earnings are mostly influenced by financial variables within the past 4 years. Apart from the significant effect of %Asset format variables for income taxes payable in our prediction model, QoQ format variables for different financial factors also contribute to the prediction greatly.

*Table 4: Feature Importance Analysis by Lagging Period*

**Feature Importance Analysis by Lagging Period**

| Lagging Period* | QoQ | YoY | %Assets | %Revenue | Total |
|---|---|---|---|---|---|
| 0 to 3 | 3 | 1 | | | 4 |
| 4 to 7 | 6 | | 6 | 2 | 14 |
| 8 to 11 | 3 | | 8 | 4 | 15 |
| 12 to 15 | 2 | | 6 | 4 | 12 |
| 16 to 19 | 4 | | | 1 | 5 |
| **Total** | **18** | **1** | **20** | **11** | **50** |

\* 0 refers to the current reporting period; 19 refers to the reporting period 5 year ago.

## 5 CONCLUSION

In this project, we have comprehensively evaluated the feasibility and suitability of adopting the Machine Learning Model (i.e., PCA + LightGBM + Hyperopt) on the forecast of company earnings, where the prediction results of our method have been thoroughly compared with both analysts' consensus estimation and traditional statistical models with logistic regression. Although at the current stage our model only outperforms logistic regression models and is unable to transcend the analysts, we do hold quite an optimistic attitude on the capability of further improvement of our models.

As mentioned, currently the analysts are largely benefited from a broader information set that contains some hidden or even insider information that can hardly be directly quantified and input into the LightGBM model. Our current machine learning model only utilizes the structured data obtained from public databases like Compustat or Thomson Reuters, but it should be well noted that the enlarged information set of the machine learning methods with the advanced techniques such as the Natural Language Processing (NLP) models may potentially exploit more non-quantitative fundamental information from various market news or messages. Future research may also explore the potential of our machine learning model by forecasting a different kind of earnings, such as a ranking of peer companies during the same testing period.

To sum up, currently, our model has already been able to serve as a favourable auxiliary tool for analysts to conduct better predictions on company fundamentals. Compared with previous traditional statistical models being widely adopted in the industry including Logistic Regression, this "PCA + LightGBM + Hyperopt" method has already achieved satisfactory advancement on both the prediction accuracy and speed. Meanwhile, we should be confident enough that there is still a vast potentiality for this model to evolve, where we do hope that in the near future, the machine learning model could generate similar or even better performances compared with professional analysts.

# Appendices

## Appendix 1: Glossary

| Name | Description |
|---|---|
| Earnings | Net income |
| Samples | Each sample has a unique combination of Company (gvkey) and Timing (datacqtr), equivalent to rows in tables |
| Variables/Dimensions/Features | Financial, macroeconomic, or return factors equivalent to columns in tables |
| Dimension reduction | Reduce the number of variables (i.e., columns) |
| Machine/LightGBM | Machine learning algorithm used to construct our model |
| Hyper-parameter | Settings for LightGBM that decide how the algorithm will construct the model |
| Classification | Problems with discrete Y options |
| Overfit | Significantly lower out-of-sample accuracy compared to in-sample accuracy |
| Dense Sample/Density | The information amount within given sample |
| Pseudocode | Informal description of the computer program |
| feature importance | The importance of each input variables in LightGBM model |

## Appendix 2: Calculation Formula

| Calculation | Formula* |
|---|---|
| QoQ | $(T_1 - T_0) \div |T_0| - 1$ |
| YoY (for X) | $(T_4 - T_0) \div |T_0| - 1$ |
| YoY (for Y – Net incomes) | $\dfrac{(T_1+T_2+T_3+T_4)-(T_0+T_{-1}+T_{-2}+T_{-3})}{|(T_0+T_{-1}+T_{-2}+T_{-3})|} - 1$ |
| %Assets | $\ln(T_0 / \text{Total Assets}_0 + 1)$ |
| %Revenue | $\ln(T_0 / \text{Total Revenue}_0 + 1)$ |
| Average of Square Residual | $\sum_{1}^{n}(T_0 - \dfrac{\sum_{1}^{p}\sum T_{-p}}{p}) \div n$  for $p$ rolling periods, and $n$ existing values |
| Price Variables | $\text{QoQ}_{\text{Company}} - \text{QoQ}_{\text{S\&P500}}$ |
| Accuracy Score | $\text{YoY}_{\text{Company}} - \text{YoY}_{\text{S\&P500}}$ |

* $T_0$ refers to the values of variables in the reporting quarter.

**Appendix 3: Data Sources**

| Purpose | Library/Files | Vendors |
| --- | --- | --- |
| Samples Selection | SAZ_MTH | CRSP |
| ETF Exclusion | FUND_HDR | CRSP |
| Financial Variables | FUNDQ | Compustat |
| Macroeconomic Variables | | Thomson Reuters |
| Return Variables | COMPD/SECM | Compustat |
| Consensus | STATSUM | I/B/E/S |

**Appendix 4: Financial Variables Description & Formats**

| Name | Description | YoY | QoQ | % Assets | % Revenue |
| --- | --- | --- | --- | --- | --- |
| acchgq | Accounting Changes - Cumulative Effect | √ | √ | | √ |
| acomincq | Accumulated Other Comprehensive Income (Loss) | √ | √ | | √ |
| acoq | Current Assets - Other - Total | √ | √ | √ | |
| actq | Current Assets - Total | √ | √ | √ | |
| ancq | Non-Current Assets - Total | √ | | √ | |
| aocipenq | Accum Other Comp Inc - Min Pension Liab Adj | √ | √ | √ | |
| aoq | Assets - Other - Total | √ | | √ | |
| apq | Account Payable/Creditors - Trade | √ | √ | √ | |
| atq | Assets - Total | √ | √ | | |
| capxy_q | Capital Expenditures | √ | √ | | √ |
| chechy_q | Cash and Cash Equivalents - Increase (Decrease) | √ | √ | | √ |
| cheq | Cash and Short-Term Investments | √ | √ | √ | |
| ciotherq | Comp Inc - Other Adj | √ | √ | | √ |
| cogsq | Cost of Goods Sold | √ | √ | | √ |
| cshopq | Total Shares Repurchased - Quarter | √ | √ | √ | |
| dcomq | Deferred Compensation | √ | | √ | |
| diladq | Dilution Adjustment | √ | √ | √ | |
| dlcq | Debt in Current Liabilities | √ | | √ | |
| dlttq | Long-Term Debt - Total | √ | | √ | |
| doq | Discontinued Operations | √ | √ | | √ |
| dpactq | Depreciation, Depletion and Amortization (Accumulated) | √ | √ | | √ |
| dpq | Depreciation and Amortization - Total | √ | √ | | √ |
| drcq | Deferred Revenue - Current | √ | √ | √ | |
| drltq | Deferred Revenue - Long-term | √ | | √ | |
| dvpq | Dividends - Preferred/Preference | √ | | √ | |
| dvy_q | Cash Dividends | √ | √ | | √ |
| esopctq | Common ESOP Obligation - Total | √ | | √ | |
| esoptq | Preferred ESOP Obligation - Total | √ | | √ | |
| fincfy_q | Financing Activities - Net Cash Flow | √ | √ | | √ |
| gdwlq | Goodwill (net) | √ | | √ | |
| intanq | Intangible Assets - Total | √ | | √ | |
| invchy_q | Inventory - Decrease (Increase) | √ | √ | | √ |
| invtq | Inventories - Total | √ | √ | √ | |
| ivchy_q | Increase in Investments | √ | √ | | √ |
| ivltq | Total Long-term Investments | √ | | √ | |
| ivncfy_q | Investing Activities - Net Cash Flow | √ | √ | | √ |
| lcoq | Current Liabilities - Other - Total | √ | √ | √ | |
| lctq | Current Liabilities - Total | √ | √ | √ | |
| lltq | Long-Term Liabilities (Total) | √ | | √ | |
| loq | Liabilities - Other | √ | | √ | |
| ltq | Liabilities - Total | √ | √ | √ | |
| mibq | Noncontrolling Interest - Redeemable - Balance Sheet | √ | | √ | |
| mibtq | Noncontrolling Interests - Total - Balance Sheet | √ | | √ | |
| miiq | Noncontrolling Interest - Income Account | √ | √ | | √ |
| niq | Net Income (Loss) | √ | √ | | √ |
| nopiq | Non-Operating Income (Expense) - Total | √ | √ | | √ |
| oancfy_q | Operating Activities - Net Cash Flow | √ | √ | | √ |
| oibdpq | Operating Income Before Depreciation - Quarterly | √ | √ | | √ |
| ppentq | Property Plant and Equipment - Total (Net) | √ | | √ | |
| pstkq | Preferred/Preference Stock (Capital) - Total | √ | | √ | |
| rcpq | Restructuring Cost Pretax | √ | √ | | √ |
| rdipq | In Process R&D | √ | √ | | √ |
| recchy_q | Accounts Receivable - Decrease (Increase) | √ | √ | | √ |
| recdq | Receivables - Estimated Doubtful | √ | √ | √ | |
| rectq | Receivables - Total | √ | √ | √ | |
| req | Retained Earnings | √ | √ | √ | |
| revtq | Revenue - Total | √ | √ | | |
| seqq | Stockholders Equity Quarterly | √ | √ | √ | |
| sivy_q | Sale of Investments | √ | √ | | √ |
| spiq | Special Items | √ | √ | | √ |
| sppivy_q | Sale of PP&E and Investments - (Gain) Loss | √ | √ | | √ |
| stkcoq | Stock Compensation Expense | √ | √ | | √ |
| tstkq | Treasury Stock - Total (All Capital) | √ | √ | √ | |
| txdbq | Deferred Taxes - Balance Sheet | √ | √ | √ | |
| txpq | Income Taxes Payable | √ | √ | √ | |
| txtq | Income Taxes - Total | √ | √ | | √ |
| wcapq | Working Capital (Balance Sheet) | √ | √ | √ | |
| xaccq | Accrued Expenses | √ | √ | | √ |
| xintq | Interest and Related Expense- Total | √ | √ | | √ |
| xiq | Extraordinary Items | √ | √ | | √ |
| xoprq | Operating Expense- Total | √ | √ | | √ |
| xrdq | Research and Development Expense | √ | √ | | √ |
| xsgaq | Selling, General and Administrative Expenses | √ | √ | | √ |

**Appendix 5: Variable Deletion**

5.1. Deleted high missing rate variables

| Name | Missing Rate | Name | Missing Rate | Name | Missing Rate |
|---|---|---|---|---|---|
| acchgq_qoq | 99.15% | dcomq_yoy | 91.77% | pstkq_yoy | 79.88% |
| ciotherq_qoq | 98.67% | diladq_yoy | 91.43% | rdipq_yoy | 79.38% |
| diladq_qoq | 98.66% | doq_yoy | 90.85% | sivy_q_yoy | 77.95% |
| doq_qoq | 98.23% | drltq_yoy | 88.12% | xiq_yoy | 74.79% |
| miiq_qoq | 98.17% | dvpq_yoy | 86.56% | cshopq_atq | 74.27% |
| rdipq_qoq | 97.49% | esopctq_yoy | 85.69% | recdq_atq | 73.68% |
| sivy_q_qoq | 94.35% | esoptq_yoy | 85.18% | txdbq_atq | 72.15% |
| xiq_qoq | 94.06% | mibq_yoy | 83.16% | rcpq_revtq | 70.97% |
| acchgq_yoy | 94.03% | mibtq_yoy | 82.97% | xaccq_revtq | 70.50% |
| ciotherq_yoy | 93.74% | miiq_yoy | 82.71% | | |

5.2 Deleted highly correlated variables

| Name | Correlated Variables | Correlation Rate | Name | Correlated Variables | Correlation Rate |
|---|---|---|---|---|---|
| aocipenq_yoy | aocipenq_qoq | 96.09% | intanq_atq | dcomq_atq | 96.05% |
| drcq_yoy | drcq_qoq | 92.39% | ivltq_atq | dcomq_atq | 92.53% |
| dvy_q_yoy | dvy_q_qoq | 91.05% | lcoq_atq | acoq_atq | 92.54% |
| txdbq_yoy | txdbq_qoq | 92.03% | loq_atq | acoq_atq | 90.63% |
| apq_atq | acoq_atq | 91.71% | mibtq_atq | mibq_atq | 94.93% |
| drltq_atq | drcq_atq | 95.74% | wcapq_atq | actq_atq | 96.08% |
| dvpq_atq | acoq_atq | 95.57% | acomincq_revtq | dcomq_atq | 95.09% |
| esoptq_atq | esopctq_atq | 97.36% | ciotherq_revtq | Aocipenq_atq | 91.78% |
| gdwlq_atq | dcomq_atq | 95.63% | rdipq_revtq | dcomq_atq | 91.38% |

*_qoq, _yoy, _atq, _revtq reprensent format YoY, QoQ, %Assets and %Revenue respectively.

**Appendix 6: Sample Elimination Result**

| Procedures | ID | Companies | Samples |
|---|---|---|---|
| Top 3000 Market Capitalization | PERMNO | 12484 | |
| Remove 1) Possible ETF; 2) Share Price Below $1; | PERMNO | 11539 | |
| No Corresponding GVKEY | GVKEY | 8325 | |
| Remove Utility and Finance | GVKEY | 6020 | |
| Remove 1) Fiscal Year End Mismatch; 2) Non-sequential Reporting History | GVKEY | 4592 | 333325 |

**Appendix 7: Optimal Rolling Period for Relevant Fill-in**

| Name | Fill-in Period | Name | Fill-in Period | Name | Fill-in Period |
|---|---|---|---|---|---|
| acchgq_revtq | 1 | dvpq_atq | 1 | oibdpq_revtq | 1 |
| acomincq_revtq | 1 | dvy_q_revtq | 1 | ppentq_atq | 1 |
| acoq_atq | 1 | esopctq_atq | 1 | pstkq_atq | 1 |
| actq_atq | 1 | esoptq_atq | 1 | rdipq_revtq | 1 |
| ancq_atq | 1 | fincfy_q_revtq | 4 | recchy_q_revtq | 1 |
| aocipenq_atq | 1 | gdwlq_atq | 1 | rectq_atq | 1 |
| aoq_atq | 1 | intanq_atq | 1 | req_atq | 1 |
| apq_atq | 1 | invchy_q_revtq | 1 | sivy_q_revtq | 1 |
| capxy_q_revtq | 1 | invtq_atq | 1 | spiq_revtq | 4 |
| chechy_q_revtq | 4 | ivchy_q_revtq | 4 | sppivy_q_revtq | 1 |
| ciotherq_revtq | 1 | ivltq_atq | 1 | stkcoq_revtq | 1 |
| cogsq_revtq | 1 | ivncfy_q_revtq | 4 | tstkq_atq | 1 |
| dcomq_atq | 1 | lcoq_atq | 1 | txpq_atq | 1 |
| diladq_atq | 1 | lctq_atq | 1 | txtq_revtq | 1 |
| dlcq_atq | 1 | lltq_atq | 1 | wcapq_atq | 1 |
| dlttq_atq | 1 | loq_atq | 1 | xintq_revtq | 1 |
| doq_revtq | 1 | mibq_atq | 1 | xiq_revtq | 1 |
| dpactq_revtq | 1 | mibtq_atq | 1 | xoprq_revtq | 1 |
| dpq_revtq | 1 | miiq_revtq | 1 | xrdq_revtq | 1 |
| drcq_atq | 1 | nopiq_revtq | 4 | xsgaq_revtq | 1 |
| drltq_atq | 1 | oancfy_q_revtq | 1 | | |

**Appendix 8: LightGBM Hyper-parameter for Optimization**

| Hyper-parameter | Testing Minimum | Testing Maximum |
|---|---|---|
| learning_rate | 0.6 | 1 |
| max_bin | 127 | 255 |
| num_leaves [1] | 50 | 200 |
| min_data_in_leaf [2] | 500 | 1400 |
| feature_fraction | 0.3 | 0.8 |
| bagging_fraction | 0.4 | 0.8 |
| bagging_freq | 2 | 8 |
| min_gain_to_split | 0.5 | 0.72 |
| lambda_l1 | 1 | 20 |
| lambda_l2 | 350 | 450 |

[1] Due attention should be paid that such parameter should not exceed $Max\_depth^2$ or the overfitting problem may occur.

[2] This is actually determined by both the size of training set and the num_leaves.

**Appendix 9: Detailed Multi-Class Prediction Results**

9.1. QoQ 3-Class Prediction

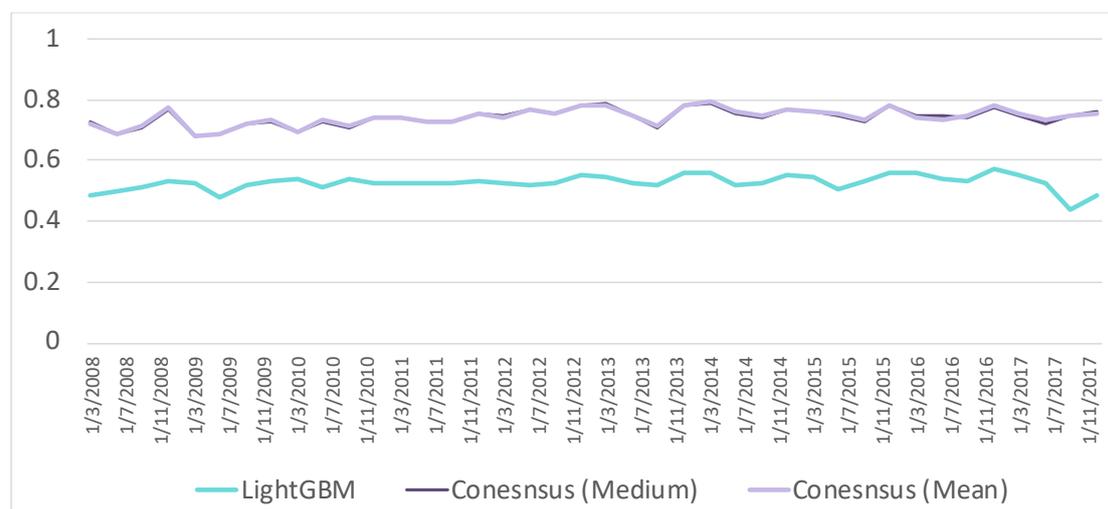

## 9.2. QoQ 6-Class Prediction

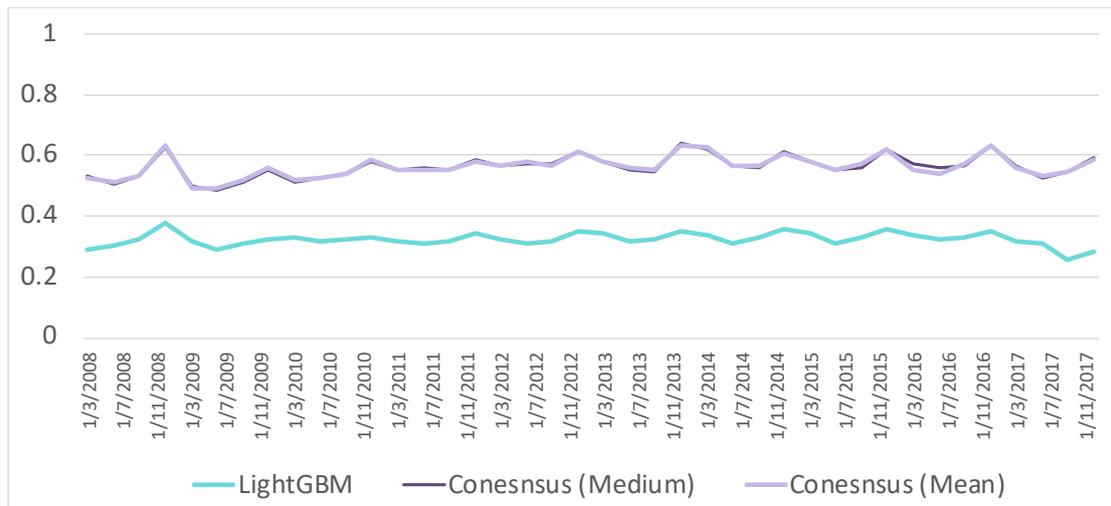

## 9.3. QoQ 9-Class Prediction

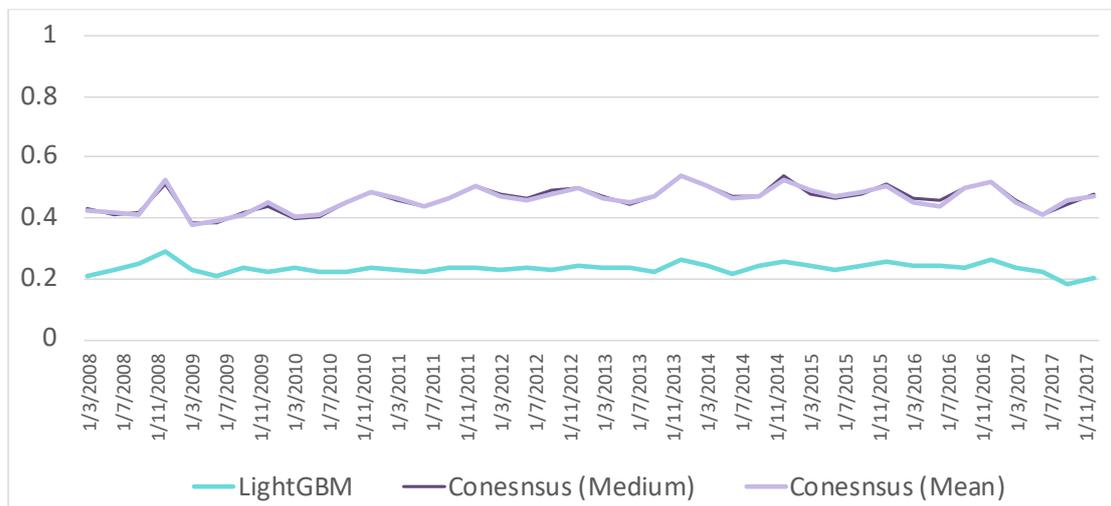

## 9.4. YoY 3-Class Prediction

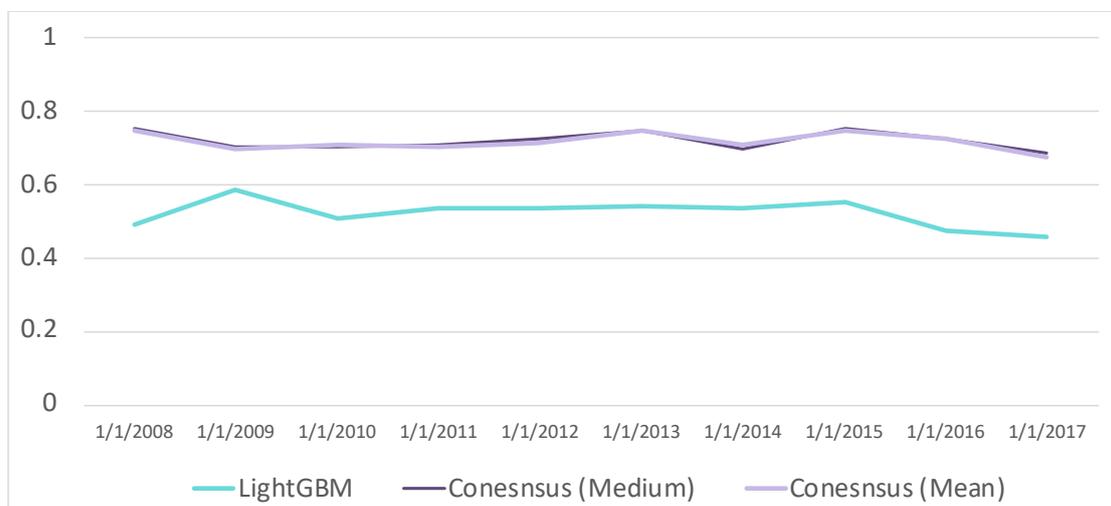

## 9.5. YoY 6-Class Prediction

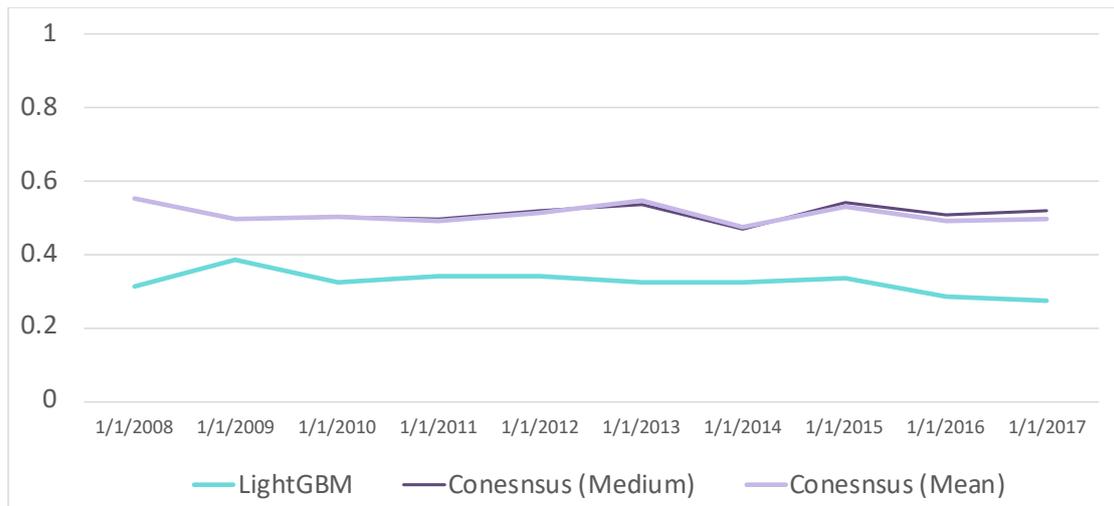

## 9.6. YoY 9-Class Prediction

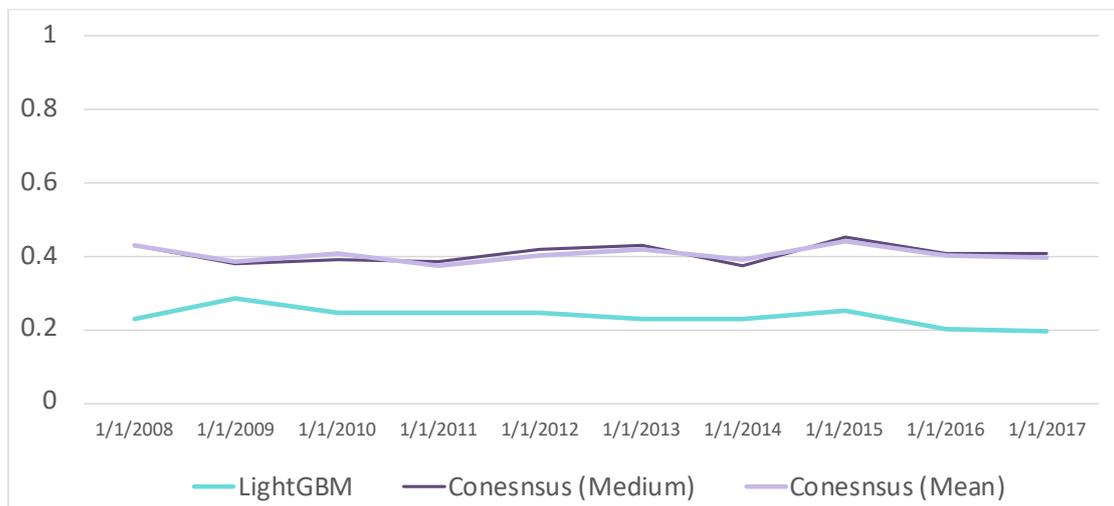

**Appendix 10: Feature Importance Result**

10.1. Most important components

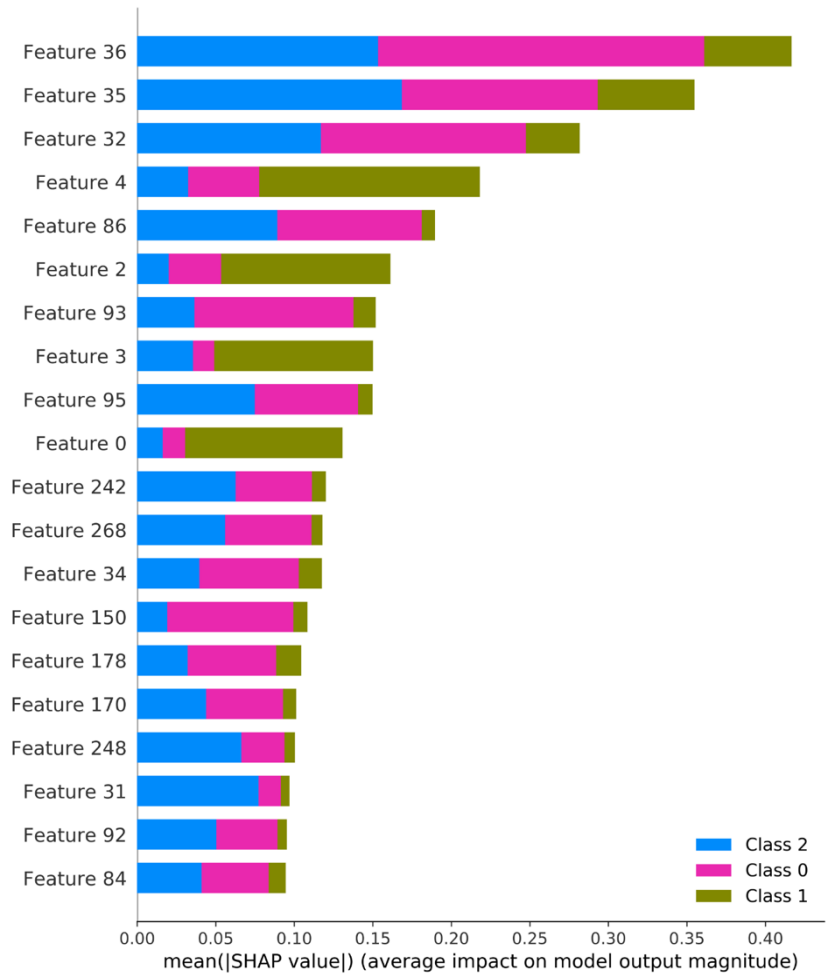

10.2. Most important variables for the top five components

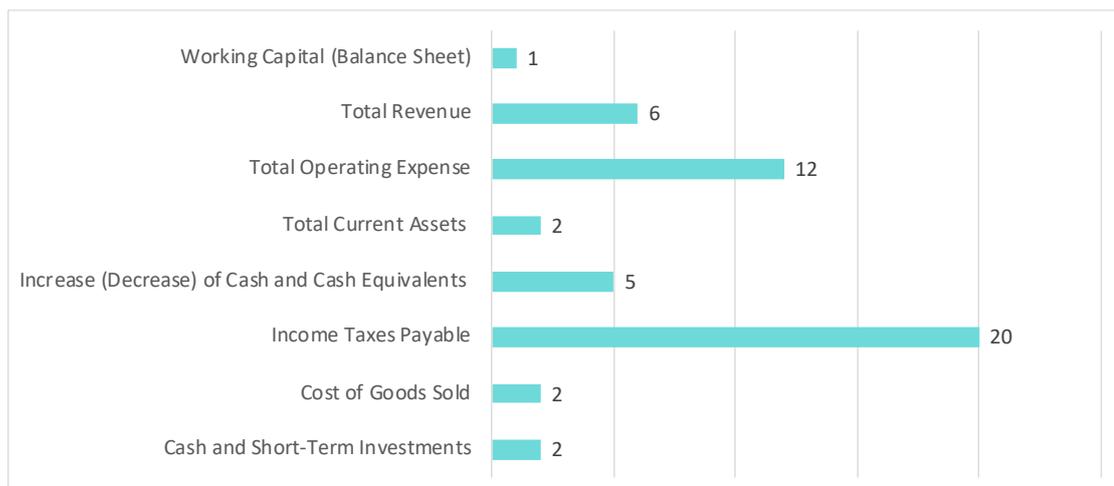

**Appendix 11: Mathematical and Computational Explanation**

11.1. Mathematical Explanation of Principal Component Analysis (PCA)

The underlying idea of PCA is to use a hyperplane to *rephrase* all of the samples, where an effective hyperplane of doing so would obviously satisfy the following criteria:

  a. Minimizing the distance of refactoring: The distance between the sample point and this hyperplane is close enough.
  b. Maximizing the separability: The projection of the sample points on this hyperplane can be separated as much as possible.

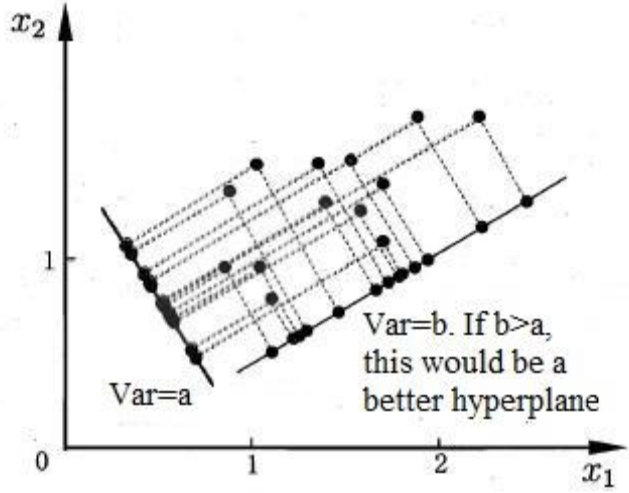

From the first property, we can easily learn that the distance between the original sample points and the projection refactored points should be minimized, which is to minimize -

$$\sum_{i=1}^{m}\left\|\sum_{j=1}^{d'} z_{ij}\boldsymbol{w}_j - \boldsymbol{x}_i\right\|^2 = \sum_{i=1}^{m} \boldsymbol{z}_i^{\mathrm{T}}\boldsymbol{z}_i - 2\sum_{i=1}^{m}\boldsymbol{z}_i^{\mathrm{T}}\mathbf{W}^{\mathrm{T}}\boldsymbol{x}_i + \text{const}$$

$$\propto -\mathrm{tr}\left(\mathbf{W}^{\mathrm{T}}\left(\sum_{i=1}^{m}\boldsymbol{x}_i\boldsymbol{x}_i^{\mathrm{T}}\right)\mathbf{W}\right)$$

Specifically, that is

$$\min_{\mathbf{W}} \quad -\mathrm{tr}(\mathbf{W}^{\mathrm{T}}\mathbf{XX}^{\mathrm{T}}\mathbf{W})$$
$$\text{s.t.} \quad \mathbf{W}^{\mathrm{T}}\mathbf{W} = \mathbf{I}$$

Besides, we can also derive the same result by maximizing the variance after projection (based on the second property), that is

$$\max_{\mathbf{W}} \quad \mathrm{tr}(\mathbf{W}^{\mathrm{T}}\mathbf{XX}^{\mathrm{T}}\mathbf{W})$$
$$\text{s.t.} \quad \mathbf{W}^{\mathrm{T}}\mathbf{W} = \mathbf{I}$$

which is obviously equivalent to the former one.

Adopting the Lagrange Multiplier Method to any of these two optimization equations, we can generate -

$$\mathbf{x}\mathbf{x}^\mathrm{T}\mathbf{w} = \lambda \mathbf{w}$$

By now, we only need to conduct the singular value decomposition to the covariance matrix $\mathbf{X}\mathbf{X}^\mathrm{T}$ and rank the eigenvalues obtained: $\lambda_1 \geq \lambda_2 \geq \cdots \geq \lambda_d$. Then, we will take the corresponding eigenvectors of the top $d'$ ones to construct $\mathbf{W}^* = (\boldsymbol{\omega_1}, \boldsymbol{\omega_2}, \dots, \boldsymbol{\omega_{d'}})$, which is the solution of PCA.

## 11.2. Pseudocode of Principal Component Analysis (PCA)

| Input | Sample Set $D = \{x_1, x_2, \dots x_m\}$; Objective dimension for the subspace $d'$ |
|---|---|
| Process | |
| 1 | Centralize/Standardize all samples: $x_i \leftarrow x_i - \frac{1}{m}\sum_{i=1}^{m} x_i$ |
| 2 | Calculate the covariance matrix for the sample $\mathbf{X}\mathbf{X}^\mathrm{T}$ |
| 3 | Conduct singular value decomposition to the covariance matrix $\mathbf{X}\mathbf{X}^\mathrm{T}$ |
| 4 | Take the corresponding eigenvectors for the largest $d'$ eigenvalues |
| Output | Projection Matrix $\mathbf{W}^* = (\boldsymbol{\omega_1}, \boldsymbol{\omega_2}, \dots, \boldsymbol{\omega_{d'}})$ |

## 11.3. Pseudocode of Decision Tree Model and Explanation

| Input | Training Set $D = \{(x_1, y_1), (x_2, y_2), \dots, (x_m, y_m)\}$; Attribute Set $A = \{a_1, a_2, \dots, a_d\}$ |
|---|---|
| Process | Function TreeGenerate($D, A$) |
| 1 | Generate the $node$ |
| 2 | if all samples in $D$ belongs to the same class $C$ then |
| 3 |   mark $node$ as a leaf node of class C; return |
| 4 | end if |
| 5 | if $A = \emptyset$ OR samples in $D$ have the same value on $A$ then |
| 6 |   mark $node$ as a leaf node, whose class label is the class that has the most samples in $D$; return |
| 7 | end if |
| 8 | Select the best partition attribute $a_*$ in $A$; |

| 9 | for each value $a_*^v$ of $a_*$ do |
|---|---|
| 10 |   generate a branch for *node*; let $D_v$ be the sample subset of $D$ that has value $a_*^v$ on $a_*$; |
| 11 |   if $D_v$ is empty then |
| 12 |     mark the branch node label as a leaf node, whose class label is the class that has the most samples in $D$; return |
| 13 |   else |
| 14 |     use TreeGenerate($D_v$, A \ $\{a_*\}$) as the branch node |
| 15 |   end if |
| 16 | end for |
| Output | A decision tree with root node - *node* |

It is easy to find that the core of a decision tree algorithm is at step 8, where the process requires us to find the optimal partition attribute $a_*$. Generally speaking, as the division process continues, we hope that the samples contained in the branch nodes of a decision tree can belong to the same class as much as possible, i.e., the 'purity' of nodes become higher and higher. Specifically, the traditional decision tree models may use *Information Entropy* (ID3 Decision Tree; Quinlan, 1986), 'Gain Ratio' (C4.5 Decision Tree; Quinlan, 1993) or the *Gini Index* (CART Decision Tree, Breiman, Friedman, Olshen & Stone, 1984) for evaluating the purity of the sample set.

Lam, M. (2004). Neural network techniques for financial performance prediction: Integrating fundamental and technical analysis. *Decision Support Systems, 37*(4), 567-581.

Lev, B., & Thiagarajan, S. R. (1993). Fundamental information analysis. *Journal of Accounting research*, 31(2), 190-215.

Lewellen, J. (2004). Predicting returns with financial ratios. *Journal of Financial Economics, 74*(2), 209-235.

Lo, A. (1994). *Data-snooping biases in financial analysis. Blending Quantitative and Traditional Equity Analysis. Charlottesville, VA: Association for Investment Management and Research.*

Lo, A., & Mackinlay, Archie Craig. (1990). *The Review of Financial Studies*, 3(3), 431-467.

Mabert, V., & Radcliffe, R. (1974). A forecasting methodology as applied to financial time series. *The Accounting Review, 49*(1), 61-75.

Myers, N., Myers, A., & Skinner, J. (2007). Earnings Momentum and Earnings Management. *Journal of Accounting, Auditing & Finance, 22*(2), 249-284.

Ndikum, P. (2020). Machine Learning Algorithms for Financial Asset Price Forecasting. *arXiv preprint arXiv:2004.01504*.

Ou, J. A., & S. H. Penman. (1989). *Financial Statement Analysis and the Prediction of Stock Returns*. Journal of Accounting and Economics 11, pp. 295-329.

Pai, P., & Lin, C. (2005). A hybrid ARIMA and support vector machines model in stock price forecasting. *Omega, 33*(6), 497-505.

Quinlan, J. R. (1986). Induction of decision trees. *Machine learning*, 1(1), 81-106.

Quinlan, J. R. (1993). *C4. 5: Programming for machine learning*. Morgan Kauffmann, 38, 48.

Richardson, S., Tuna, İ., & Wysocki, P. (2010). Accounting anomalies and fundamental analysis: A review of recent research advances. *Journal of Accounting and Economics, 50*(2), 410-454.

Schölkopf, B., Smola, A., & Müller, K. R. (1998). Nonlinear component analysis as a kernel eigenvalue problem. *Neural computation*, 10(5), 1299-1319.

Securities and Exchange Commission (SEC). (2003, September 26). Acceleration of Periodic Report Filing Dates and Disclosure Concerning Website Access to Reports. Retrieved from: https://www.sec.gov/rules/final/33-8128.htm

Sun, X., Liu, M., & Sima, Z. (2018). A novel cryptocurrency price trend forecasting